\documentclass[game]{apjrnl}

\usepackage{amsmath}
\usepackage{amssymb}
\newtheorem{proposition}{Proposition}
\newtheorem{theorem}{Theorem}
\newtheorem{lemma}{Lemma}
\newtheorem{corollary}{Corollary}

\newcommand{\argmax}{\mathop{\mbox{argmax}}}
\newcommand{\type}{\tau}
\newcommand{\Type}{\mathcal{T}}

\newcommand{\comment}[1]{}
\newcommand{\signal}{s}
\newcommand{\Signal}{\mathcal{S}}
\newcommand{\bcmax}{\kappa}

\begin{document}

\title{ Bidding Clubs in First-Price Auctions }
\author{Kevin Leyton-Brown, Yoav Shoham and Moshe Tennenholtz \\ Department of Computer Science\\ Stanford University \\ Stanford, CA 94305 \\ Email: \{kevinlb;shoham;moshe\}@cs.stanford.edu }

\maketitle

\begin{abstract}
\comment{Coordination and cooperation among self-interested agents is a problem of fundamental importance to researchers in AI and in multi-agent systems. Motivated by the application of multi-agent systems to electronic commerce, } 

We introduce a class of mechanisms, called {\em bidding clubs}, that allow agents to coordinate their bidding in auctions. Bidding clubs invite a set of agents to join, and each invited agent freely chooses whether to accept the invitation or whether to participate independently in the auction. Agents who join a bidding club first conduct a ``pre-auction'' within the club; depending on the outcome of the pre-auction some subset of the members of the club bid in the primary auction in a prescribed way. We model this setting as a Bayesian game, including agents' choices of whether or not to accept a bidding club's invitation.  After describing this general setting, we examine the specific case of bidding clubs for first-price auctions.  We show the existence of a Bayes-Nash equilibrium where agents choose to participate in bidding clubs when invited and truthfully declare their valuations to the coordinator.  Furthermore, we show that the existence of bidding clubs benefits all agents (including both agent inside and outside of a bidding club) in several different senses.

\comment{
}

\end{abstract}

\section{Introduction}

The advent of internet markets has spurred new interest in auctions.  
Most work in both economics and computer science has concentrated on the design of auction protocols from the seller's perspective, and in particular on optimal (i.e., revenue maximizing) auction design. In this paper we present a class of systems to assist sets of bidders, {\em bidding clubs.} The idea is similar to the idea behind ``buyer clubs'' on the Internet (e.g., $\underline{\mathrm{www.mobshop.com}}$): to aggregate the market power of individual bidders. Buyer clubs work when buyers' interests are perfectly aligned; the more buyers join in a purchase the lower the price for everyone.  In auctions held on the internet it is relatively easy for multiple agents to cooperate, hiding behind a single auction participant. Intuitively, these bidders can gain by causing others to lower their bids in the case of a first-price auction or by possibly removing the second-highest bidder in the case of a second-price auction.  However, the situation in auctions is not as simple as in buyer clubs, because while bidders can gain by sharing information, the competitive nature of auctions means that bidders' interests are not aligned.  Thus there is a complex strategic relationship among bidders in a bidding club, and bidding club rules must be designed accordingly.

\subsection{Related Work}\label{background}

While there is relative scarcity of previous work on bidder-centric mechanisms, certainly our work has not been carried out in a vacuum. Below we discuss the most relevant previous work and its relation to ours. This work all comes under the umbrella of {\em collusion} in auctions, a negative term still reflecting a seller-oriented perspective. We adopt a more neutral stance towards such bidder activities and thus use the term {\em bidding clubs} rather than the terms {\em bidding rings} and \emph{cartels} that have been used in the past. However, the technical development is not impacted by such subtle differences in moral attitude.

\subsubsection{Collusion in Second-Price Auctions}

One of the first formal papers to consider collusion in second-price auctions was written by Graham and Marshall \cite{GrahamMarshall}. This paper introduces a knockout procedure: agents announce their bids in a pre-auction; only the highest bidder goes to the auction but this bidder must pay a ``ring center''
the amount of his gain relative to the case where there was no collusion. The ring center pays each agent in advance; the amount of this payment is calculated so that the ring center will budget-balance \emph{ex-ante}, before knowing the agents' valuations. 

Graham and Marshall's work has been extended to deal with variations in the knockout procedure, differential payments, and relations to the Shapley value \cite{GrahamMarshallRichard}.  The case where only some of the agents are part of the cartel is discussed by Mailath and Zemsky \cite{MailathZemsky}.  Ungern and Sternberg \cite{Ungern-Sternberg} discuss collusion in second-price auctions where the designated winner of a cartel is not the agent with the highest valuation. Finally, although this fact is not presented in any existing work of which we are aware, it is also easy to extend Graham and Marshall's protocol to handle an environment where multiple cartels may operate in the same auction alongside independent bidders.  

Overall, a much richer body of work deals with second-price auctions than with first-price auctions.  This is possibly explained by the fact that since second-price auctions give rise to dominant strategies, it is possible to study collusion in many settings related to these auctions without performing strategic equilibrium analysis.  

\subsubsection{Collusion in First-Price Auctions}

The key exception to the scarcity of formal work on first-price auctions is a very influential paper by McAfee and McMillan \cite{McAfeeMcMillan92}. It is the closest in the literature to our work, and indeed we have borrowed some modelling elements from it.  Several sections of their paper, including the discussion of enforcement and the argument for independent private values as a model of agents' valuations, are directly applicable to our paper.  However, the setting introduced in their work assumes that a fixed number of agents participate in the auction and that all agents are part of a single cartel that coordinates its behavior in the auction.  The authors show optimal collusion protocols for ``weak'' cartels (in which transfers between agents are not permitted: all bidders bid the reserve price, using the auctioneer's tie-breaking rule to randomly select a winner) and for ``strong'' cartels (the cartel holds a pre-auction, the winner of which bids the reserve price in the main auction while all other bidders sit out; the winner distributes some of his gains to other cartel members through side payments). A small part of the paper deals with the case where in addition to the single cartel there are also additional agents. However, results are shown only for two cases: (1) when non-cartel members bid without taking the existence of a cartel into account and (2) when each agent $i$ has valuation $v_i \in \{0,1\}$. The authors explain that they do not attempt to deal with general strategic behavior in the case where the cartel consists of only a subset of the agents; furthermore, they do not consider the case where multiple cartels can operate in the same auction. Finally, a brief presentation of ``cartel-formation games'' is related to our discussion of agents' decision of whether or not to accept an invitation to join a bidding club.

\subsubsection{Other Work on Collusion}

Less formal discussion of collusion in auctions can be found in a wide variety of papers. For example, a survey paper that discusses mechanisms that are likely to facilitate collusion in auctions, as well as methods for the detection of such schemes, can be found in \cite{HendricksPorter}. A discussion and comparison of the stability of rings associated with classical auctions can be found in \cite{Robinsoncoll}. That paper concentrates on the case where the valuations of agents in the cartel are honestly reported. 

Collusion is also discussed in other settings. For example, the literature discusses collusion that aims to influence purchaser behavior in a repeated procurement setting (see \cite{FeinsteinBlockNold}), and in the context of general Bertrand or Cournot competition (see \cite {CramtonPalfrey}).

We should also mention that in an earlier paper we have anticipated some of the results reported here. Specifically, in \cite{biddingclubs} we considered bidding clubs under the assumptions that only a single bidding club exists, and that bidders who were not invited to join the club are not aware of the possibility that a bidding club might exist.  The current paper is an extension and generalization of that earlier work.

\subsection{Distinguishing Features of our Model}\label{tech-bg-bc}

Our goal in this work is to study cooperation between self-interested bidders in a rich model that captures many of the characteristics of auctions on the internet.  This leads to many differences between our model and models proposed in the work surveyed above (particularly \cite{GrahamMarshallRichard} and \cite{McAfeeMcMillan92}).  In particular, we argue that a model of an internet auction setting that includes bidding clubs should include the following features:

\begin{enumerate}
\item The number of bidders is stochastic.
\item There is no minimum number of bidders in a bidding club (i.e., bidding clubs are not required to contain all bidders).\footnote{For technical reasons we will have to assume that there is a finite \emph{maximum} number of bidders in each bidding club; however, this maximum may be any integer greater than or equal to two.}
\item There is no limit to the number of bidding clubs in a single auction.
\item Club members and independent bidders behave strategically, acting according to correct beliefs about this complex environment. 
\end{enumerate}

The first feature above is crucial.  In many real-world internet auctions, bidders are not aware of the number of other agents in the economic environment.  A bidding club that drops one or more interested bidders is thus undetectable to other bidders in an internet auction. An economic environment with a fixed number of bidders would not model this uncertainty, as the number of interested bidders would be common knowledge among all bidders regardless of the number of bids received in the auction. For this reason, we consider economic environments where the number of bidders is chosen at random. We make use of a model of auctions with stochastic numbers of participants which is due to McAfee and McMillan \cite{McAfeeRandom}; we also refer to equilibrium analysis of this model by Harstad, Kagel and Levin \cite{HarstadKagelLevin}.

\subsection{Bidding Clubs at a Glance}\label{bc-glance}

Roughly speaking, a scenario with bidding clubs has the following structure:

\begin{enumerate}
\item Given a primary auction;
\item Given a set of bidders in that auction, drawn randomly from a set of potential bidders;
\item Given a partition of bidders into disjoint {\em clubs,} each of which can be the redundant singleton club;
\item Each bidder chooses whether to bid in the primary auction directly or through his club (it is assumed that this choice is strictly enforceable).  In the latter case, the bidder declares his valuation to the club coordinator; 
\item Based on the bidders' choices and declarations each club bids in the primary auction, as do both the bidders who elected not to join their respective clubs and the singleton bidders. 
\item Each (non-singleton) club bids according to pre-specified, commonly known rules. These rules also specify internal allocations and possible monetary transfers among club members upon the conclusion of the primary auction.
\end{enumerate}

To make bidding clubs a more realistic model of collusion in internet auctions, we restrict bidding club protocols in the following ways:

\begin{enumerate}
\item Participation in bidding clubs requires an invitation, but bidders must be free to decline this invitation without (direct) penalty.  In this way we include the choice to collude as one of agents' strategic decisions, rather than starting from the assumption that agents will collude.
\item Bidding club coordinators must make money on expectation, and must never lose money. \label{makemoney}  This ensures that third-parties have incentive to run bidding club coordinators.  Note that this requirement is not satisfied by a \cite{GrahamMarshallRichard}-type result, in which bidding clubs (or, in their parlance, cartels) are budget balanced {\em ex ante}, but may lose money in individual auctions.
\item The bidding club protocol must give rise to an equilibrium where all invited agents choose to participate, even when the bidding club operates in a single auction as opposed to a sequence of auctions.  This means that agents can not be induced to collude in a given auction by the threat of being denied future opportunities to collude.
\end{enumerate}

\subsection{Overview}


This paper consists of two parts.  First, sections \ref{auctionmodel} through \ref{asymm-section} present relevant background that does not directly concern cooperation between bidders. In section \ref{auctionmodel} we give a formal model of an auction with a stochastic number of participants based on the model in \cite{McAfeeRandom}.  We set up an economic environment in which a finite number of agents is chosen at random from an infinite set of potential agents.  We also give a general model of auction mechanisms based on \cite{MonTenaij}, and define symmetric Bayes-Nash equilibria for the resulting Bayesian game.  In section \ref{first-price-intro} we consider different variations on the first-price auction mechanism.  We begin with classical first-price auctions, in which the number of bidders is common knowledge, and then consider first-price auctions in the economic environment from section \ref{auctionmodel}, where the number of bidders is drawn from a known distribution.  Combining results from both auction types, we present first-price auctions with participation revelation: auctions in which the number of bidders is stochastic, but the auctioneer announces the number of participants before taking bids.  This is the auction mechanism upon which we will base our bidding club protocol for first-price auctions.  Finally, section \ref{asymm-section} makes use of the revelation principle to show a class of auction mechanisms in which bidders are subject to different payment rules and may have different private information (in addition to their valuations), yet all bid truthfully.  We think that this result is interesting in its own right, and certainly it is applicable to settings other than collusion; however, it is also necessary to the proof of the main theorem in section \ref{first-price}.

The second part of our paper is concerned explicitly with bidding clubs, using material from the first part to present a general model of bidding clubs and then a bidding club protocol for first-price auctions.  First, section \ref{biddingclubs} expands the economic environment from section \ref{auctionmodel} to include the following novel features:

\begin{itemize}
\item A finite set of bidding clubs is selected from an infinite set of potential bidding clubs.
\item A finite set of agents is selected to participate in the auction, from an infinite set of potential agents.  Some agents are associated with bidding clubs, and the whole procedure is carried out in such a way that no agent can gain information about the total number of agents in the economic environment from the fact of his own selection.
\item The space of agent types is expanded to include both an agent's valuation, and the number of agents present in that agent's bidding club (equal to one if the agent does not belong to a bidding club).
\end{itemize}

We introduce notation to describe each agent's beliefs about the number of agents in the economic environment, conditioned on that agent's private information.  We also augment the auction mechanism from section \ref{auctionmodel} to describe additional strategic choices available to agents invited to bidding clubs.  In section \ref{first-price} we examine bidding club protocols for first-price auctions.  We begin with two assumptions on the distribution of agent valuations: the first related to continuity of the distribution, and the second to monotonicity of equilibrium bids.  After a technical lemma relating equilibrium bids in auctions with stochastic numbers of participants under different distributions, we give a bidding club protocol for first-price auctions with participation revelation.  Our main technical results follow:

\begin{itemize}
\item We show that it is an equilibrium for agents to accept invitations to join bidding clubs when invited and to disclose their true valuations to their bidding club's coordinator. Under the same equilibrium, singleton agents bid as they would in an auction with a stochastic number of participants in an economic environment without bidding clubs, in which the distribution over the number of participants is the same as in the bidding clubs setting.
\item In equilibrium each agent is better off as a result of his own club (that is, his expected payoff is higher than would have been the case if his club never existed, but other clubs---if any---still did exist). 
\item In equilibrium each club increases all non-members' expected payoffs, as compared to equilibrium in the case where all club members participated in the auction as singleton bidders, but all other clubs---if any---still existed.
\item In equilibrium each agent's expected payoff is identical to the case in which no clubs exist; note that since clubs make money on expectation, if clubs are willing to make money (or break even) only on expectation, they could distribute some of their {\em ex ante} expected profits among the club members, ensuring that all bidders gain on expectation.  
\end{itemize}

Finally, sections \ref{discussion} and \ref{conclusion} consist of discussion and conclusions. We touch on questions of trustworthiness of coordinators, legality of bidding clubs and steps an auctioneer could take to disrupt the operation of bidding clubs in her auction.

%
%
%

\section{Auction Model}\label{auctionmodel}


In this section we provide a (non-controversial) auction model, meant to capture an internet auction setting such as eBay.  Of course, this model is applicable to many other auctions as well.  Auctions may be seen as consisting of an economic environment plus an auction mechanism which together define a Bayesian game. First, our economic environment consists of a stochastic number of agents, each of which has private information about the number of participants in the auction and knows the distribution from which others' types are drawn.  This section draws heavily on work by McAfee and McMillan \cite{McAfeeRandom} on auctions with a stochastic number of participants. Second, the game includes an auction mechanism in which the agents participate; this section is based on \cite{MonTenaij}.  After defining these elements, we give a formal definition of the Bayesian game.

\subsection{The Economic Environment}\label{econenviron}

An economic environment $E$ consists of a finite set of agents who have non-negative valuations for a good at auction, and a distinguished agent $0$---the seller or center. The set of agents is selected by an exogenous process, and each agent is unaware of the total number of agents participating in the economic environment. Following \cite{McAfeeRandom}, let the set of agents who may participate in the economic environment be $\mathcal{A} \equiv \mathbb{N}$. Let $\beta_A$ represent the probability that a finite set $A \subset \mathcal{A}$ is the set of agents.  The probability that $n$ agents\footnote{When we say that $n$ agents participate in the auction we do not count the distinguished agent $0$, who is always present.} will participate in the auction is $\gamma_A(n) = \sum_{A, |A| = n} \beta_A$. All agents know the probability distribution $\beta_A$. Once an agent $k$ is selected, he updates his probability of the number of agents present as:

\begin{equation}
p_n^k = \frac{\sum_{A, |A| = n, k \in A} \beta_A}{\sum_{A, k \in A} \beta_A}.
\end{equation}

We deviate from the model in \cite{McAfeeRandom} by adding the assumption 
that it is common knowledge that all bidders are equally likely to be chosen.  Hence $p_n^k$ is the same for all $k$; we will hereafter refer only to $p_n$.  
Finally, we assume that $\gamma_A(0) = \gamma_A(1) = 0$; at least two agents will participate in the auction.

Let $\mathcal{T}$ be the set of possible agent types. The type $\type_i \in \Type$ of agent $i$ is the tuple $(v_i, \signal_i) \in V \times \Signal$.  $v_i$ denotes an agent's valuation: his maximal willingness to pay for the good offered by the center.  We assume that $v_i$ represents a purely private valuation for the good, and that $v_i$ is selected independently from the other $v_j$'s of other agents from a known distribution,
%
%
$F$, having density function $f$.  By $\signal_i$ we denote agent $i$'s signal: his private information about the number of agents in the auction.  In this section we will consider the simple case where $\Signal = \{\varnothing\}$: it is common knowledge that all agents receive the null signal, and hence gain no additional information about the number of agents. Note, however, that the economic environment itself is always common knowledge, and so agents always have some information about the number of agents even when they receive the null signal. We will consider more complex signals in section \ref{biddingclubs}. We will use the notation $p_n^{\type_i}$ to denote the probability that agent $i$ assigns to there being $n$ agents in the auction, conditioned on his type $\type_i$.  Throughout the paper we will use uppercase $P$ to denote the whole probability distribution as compared to the probability of a particular number of agents which we have denoted by lowercase $p$; in this case we denote the whole distribution conditioned on $i$'s type as $P^{\type_i}$.

The utility function of agent $i$, $u_i:\mathbb{R} \rightarrow \mathbb{R}$ is linear, normalized with $u_i(0)=0$.  The utility of agent $i$ (having valuation $v_i$) when asked to pay $t$ is $v_i-t$ if $i$ is allocated a good, and it is $0$ otherwise.  Thus, we assume that there are no externalities in agents' valuations and that agents are risk-neutral.

\subsection{The Auction Mechanism}\label{auctionmech}


We denote the possible allocations of the good to the agents by $\Pi$.  An auction mechanism is a tuple $(\mathcal{M},g,t)$, where:


\begin{itemize}
\item $\mathcal{M}$ is the set of possible messages an agent may send.
\item $g:\mathcal{M}^n \rightarrow \Delta(\Pi)$ is an allocation function where  $\Delta(\Pi)$  is the tuple of distribution functions over $\Pi$ (e.g., the allocation may include random elements).
\item $t=(t_1,t_2,\ldots,t_n)$, $t_i: \mathcal{M}^n \times \Pi \rightarrow \mathbb{R}$ is the (monetary)  transfer function for agent $i$.
\end{itemize}

Notice that $n$ is a parameter. Technically, an auction mechanism defines $g$ and $t$ for any number of participants, and can be therefore considered as a set of tuples (one for each number of agents).

Given the above, the dynamics of an auction mechanism can be described as follows:

\begin{itemize}
\item Each agent $i$ sends a message $\mu_i$ to the center.  We denote the set of messages received by the center as $\mu$.
\item The center conducts a lottery according to the distribution $g(\mu)$, and selects the allocation $\pi$.
\item Agent $i$ gets $\pi_i$, and is required to transfer $t_i(\mu,\pi)$ to the center.
\item The utility of $i$ is $v_i- t_i(\mu,\pi)$ if he is assigned a good, and it is $-t_i(\mu,\pi)$ otherwise.
\end{itemize}


\subsection{The Bayesian Game}\label{bayesgame}

The auction mechanism $(\mathcal{M},g,t)$, in conjunction with the economic environment $E$, defines a Bayesian game. We will use the following definitions and notation.  A strategy $b_i:\Type \rightarrow \mathcal{M}$ for agent $i$ is a mapping from his type $\type_i$ to a message $\mu_i$. This may be the null message, which means that he has elected not to participate in the auction.  $\Sigma$ denotes the set of possible strategies, i.e., the set of functions from types to messages in $\mathcal{M}$. Each agent's type is that agent's private information, but the whole setting is common knowledge.
%
%

For notational simplicity we only define symmetric equilibria, where all agents bid the same function of their type, as this is sufficient for our purposes in this paper.  A more general definition would proceed along the same lines.
By $L_i(\type_i,b_i,b^{j-1})$ we denote agent $i$'s \emph{ex post} expected utility given that his type is $\type_i$, he follows the strategy $b_i$ and all other agents use the strategy $b$, in the case that there are a total of $j$ agents. The strategy profile $b^n \in\Sigma^n$ is a symmetric equilibrium if and only if:

\begin{equation}
\forall i\in A, \forall \type_i \in \Type, b \in \argmax_{b_i\in\Sigma} \sum_{j=2}^{\infty} p_j^{\type_i} L_i(\type_i,b_i,b^{j-1})
\end{equation}





\section{First-Price Auctions}\label{first-price-intro}

In this section we discuss several different variants of the first-price auction.  First we describe classical first-price auctions, in which a fixed number of participants belong to the economic environment, and hence the number of bidders is common knowledge.  Next we consider first-price auctions with a stochastic number of participants, where the number of bidders in the economic environment is drawn from a known distribution.  Using the previous two settings, we present first-price auctions with participation revelation, where the number of agents is chosen stochastically, but the auctioneer announces the number of agents who have registered in the auction before taking bids.  This last type of first-price auction is the one we will consider in our discussion of bidding clubs in section \ref{first-price}.

\subsection{Classical first-price auctions}\label{classicalfirstprice}

In a classical first-price auction, each participant submits a bid in a sealed envelope. The agent with the highest bid wins the good and pays the amount of his bid, and all other participants pay nothing. In the case of a tie, the winner of the auction is selected uniformly at random from the bidders who tied for the highest bid. (Note, however, that when $F$ is continuous and has no atoms the probability of two bidders having the same type is $0$; ties will therefore occur with probability $0$ if bidders follow an equilibrium in which they all bid a strictly monotonically-increasing function of their valuations.) The equilibrium analysis of first-price auctions is quite standard: 

\begin{proposition}If valuations are selected independently according to the uniform distribution on $[0,1]$ then it is a symmetric equilibrium for each agent $i$ to follow the strategy:

\begin{equation*}
b(v_i)=\frac{n-1}{n} \, v_i.
\end{equation*}  
\end{proposition}

Using classical equilibrium analysis (e.g., following Riley and Samuelson \cite{RileySamuelson}) it is possible to show how classical first-price auctions can be generalized to an arbitrary continuous distribution $F$.

\begin{proposition}\label{prop-first-fixed}If valuations are selected from a continuous distribution $F$ then it is a symmetric equilibrium for each agent $i$ to follow the strategy:

\begin{equation*}
b(v_i)=v_i-F(v_i)^{-{{(n-1)}}}\int_0^{v_i} {F(u)^{n-1}} du.
\end{equation*}.
\end{proposition}

In both cases, observe that although $n$ is a free variable, $n$ is not a parameter of the strategy; the same is true of the distribution $F$.  Agents deduce this information from their full knowledge of the economic environment.  It is useful, however, to have notation specifying the amount of the equilibrium bid as a function of both $v$ and $n$.  We write

\begin{equation}\label{be-n}
b^e(v_i,n)=v_i-F(v_i)^{-{{(n-1)}}}\int_0^{v_i} {F(u)^{n-1}} du. 
\end{equation}

\subsection{First-price auctions with a stochastic number of bidders}\label{first-price-stochastic}

In the economic environment described in section \ref{econenviron} the number of agents is not a constant; rather, it is chosen stochastically from a known probability distribution.  An equilibrium for this setting was demonstrated by Harstad, Kagel and Levin \cite{HarstadKagelLevin}:

\begin{proposition}\label{propstochastic}If valuations are selected from a continuous distribution $F$ and the number of bidders is selected from the distribution $P$ then it is a symmetric equilibrium for each agent $i$ to follow the strategy:

\begin{equation*}
b(v_i) = \sum_{j=2}^{\infty}p_j b^e(v_i,j)
\end{equation*}
\end{proposition}

Observe that $b^e(v_i,j)$ is the amount of the equilibrium bid for a bidder with valuation $v_i$ in a setting with $j$ bidders as described in section \ref{classicalfirstprice} above.  $P$ is deduced from the economic environment.\footnote{Recall that $P$ is a set: $p_j \in P$ for all $j \geq 0$, where $p_j$ denotes the probability that the economic environment contains exactly $j$ agents.}  We overload our previous notation for the equilibrium bid, this time as a function of the agent's valuation and the probability distribution $P$.  Thus we write:

\begin{equation}\label{stochasticbid}
b^e(v_i,P) = \sum_{j=2}^{\infty}p_j b^e(v_i,j)
\end{equation}

We will make frequent use of this function throughout the paper.  An important note is that it describes the equilibrium bid in the situation where the economic environment is such that the number of agents is chosen by $P$ and where all agents receive the null signal.

\subsection{First-price auctions with participation revelation}\label{participationrevelation}

In some first-price auctions (e.g., auctions held on the internet), bidders participate in an economic environment where the number of bidders in the auction is not common knowledge.  However, this can be helpful information for bidders. One obvious way of addressing this problem is to introduce a two-phase mechanism with revelation of the number of participants between the stages.  Specifically, a first-price auction with participation revelation is as follows:

\begin{enumerate}
\item Agents indicate their intention to bid in the auction.
\item The auctioneer announces $n$, the number of agents who registered in the first phase.
\item Agents submit bids to the auctioneer. The auctioneer will only accept bids from agents who registered in the first phase.
\item The agent who submitted the highest bid is awarded the good for the amount of his bid; all other agents are made to pay 0.
\end{enumerate}


It is unsurprising that, although a first-price auction with participation revelation may have a stochastic number of participants,

\begin{proposition}\label{prop-participationrevelation}
There exists an equilibrium of the first-price auction with participation revelation where every agent $i$ indicates the intention to participate and bids according to $b^e(v_i,n)$.
\end{proposition}

\begin{proof}
Agents are always better off participating in first-price auctions as long as there is no participation fee.  The only way of participating is to declare the intention to participate in the first phase of the auction.  Thus the number of agents announced by the auctioneer is equal to the total number of agents in the economic environment. From proposition \ref{prop-first-fixed} it is best for agent $i$ to bid $b^e(v_i,n)$ when it is common knowledge that the number of agents in the economic environment is $n$. That is exactly the case under our mechanism.
\end{proof}


In section \ref{first-price} we will be concerned with first-price auctions with information revelation, but we will show an equilibrium in which the number of agents registering in the first phase is smaller than the total number of agents participating in the auction, because some bidders with low valuations drop out as part of a collusive agreement.  The auctioneer's declaration acts as a signal about the total number of bidders, but individual agents will still be uncertain about the total number of opponents they face.

\section{Truthful Equilibria in Asymmetric Mechanisms}\label{asymm-section}

In this section we describe a particular class of auction mechanisms that are asymmetric in the sense that every agent is subject to the same allocation rule but to a potentially different payment rule, and furthermore that agents may receive different signals.  It will be helpful for the proof of our main theorem in section \ref{first-price} to show that a truth-revealing equilibrium exists in such auctions under the following two conditions:

\begin{enumerate}
\item The auction allocates the good to the agent who submits the highest bid.
\item Consider the auction $M_i$ in which \emph{all} agents are subject to agent $i$'s payment rule and the above allocation rule, and where (hypothetically) \emph{all} agents receive the signal $\signal_i$.\footnote{That is, for every agent $j$ in the real auction, we create an agent $k$ in the hypothetical auction $M_i$ having type $\type_k = (v_j,\signal_i)$.}  Truth-revelation is a symmetric equilibrium in $M_i$. 
\end{enumerate}

Observe that the second condition above is less restrictive than it may appear.  From the revelation principle we can see that for every auction with a symmetric equilibrium there is a corresponding auction in which truth-revealing is an equilibrium that gives rise to the same allocation and the same payments for all agents. $M_i$ can thus be seen as a revelation mechanism for some other auction that has a symmetric equilibrium.

More formally, given a good $g$, let $\textbf{M}$ represent a set of auctions $\{M_1, \ldots,$ $M_n\}$ which all allocate the good to the agent who submits the highest bid, and which are all truth-revealing direct mechanisms for $n$ risk-neutral agents with independent private valuations drawn from the same distribution. We now define another auction $\bar{M}$: 

\begin{enumerate} 
\item Each agent $i$ sends a message $\mu_i$ to the center.  
\item The center allocates the good to the agent $i$ with $\mu_i \in \max_j \mu_j$.  If multiple agents submit the highest message, the tie is broken in some arbitrary way.
\item Agent $i$ is made to transfer $t_i(\mu,\pi)$ to the center.\footnote{Of course, this transfer can be either positive or negative.}  The transfer function $t_i$ is taken from $M_i \in {\textbf{M}}$.
\end{enumerate}

We can now show: 

\begin{lemma}\label{lemma} 
Truth-revelation is an equilibrium of $\bar{M}$.
\end{lemma} 

\begin{proof}
The payoff of agent $i$ is uniquely determined by the allocation rule, the transfer function $t_i$, and all agents' strategies. Assume that the other agents are truth revealing, then the other agents' behavior, the allocation rule, and agent $i$'s payment rule are all identical in $\bar{M}$ and $M_i$. Since truth-revelation is an equilibrium in $M_i$, truth-revelation is agent $i$'s best response in $\bar{M}$. 
\end{proof}

\paragraph{Example.} Consider an auction for a single good $g$, where eight agents bid for the good.  The agents' valuations are IPV, IID from a known distribution $F$, and the agents are risk-averse.  Let $M_1$ be a revelation mechanism for a first-price auction: i.e., agents declare their valuations, and the winner is charged $b^e(v,8)$.  In an economic environment consisting of eight agents with IPV valuations from $F$ it is an equilibrium of $M_1$ for agents to truthfully declare their valuations to the center.  Let $M_2$ be a second-price auction; truthful declaration is a weakly dominant strategy under this auction type.  Both $M_1$ and $M_2$ allocate the good to the agent with the highest declaration, and so these auctions meet the conditions given at the beginning of the section. Now consider an auction $\bar{M}$ where odd-numbered agents are subject to the payment rule from $M_1$, and even-numbered agents are subject to the payment rule from $M_2$.  By lemma \ref{lemma}, truth-revelation is an equilibrium of $\bar{M}$.  
There are other differences between payment rules that can 
cause agents' expected utilities to differ: for example, lemma \ref{lemma} would still hold if $M_2$ gave each agent an additional payment of $\$10$ for participating in the auction.

The next corollary, which follows directly from the lemma, compares a single agent's expected utility under two different auctions $\bar{M}$ and $\bar{M}'$, which implement different payment rules.  We will need this result for our proof of theorem \ref{theorem-firstprice}.

\begin{corollary}\label{asymm-cor}
Consider two auctions $\bar{M}$ and $\bar{M}'$, defined as above, which both implement the same transfer function for agent $i$.  Agent $i$'s expected utility is the same in both $\bar{M}$ and $\bar{M}'$.
\end{corollary}

\begin{proof}
The payoff of agent $i$ is uniquely determined by the allocation rule, its transfer function, and all agents' strategies. Both $\bar{M}$ and $\bar{M}'$ have the same allocation rule. Lemma \ref{lemma} tells us that truth revelation is a best response for all agents in both $\bar{M}$ and $\bar{M}'$, so all agents' strategies are identical in the two auctions.  In general, agents may not receive the same expected utility from $\bar{M}$ and $\bar{M}'$.  However, since $i$ has the same transfer function in both auctions, $i$'s expected utility in $\bar{M}$ is equal to his expected utility in $\bar{M}'$.
\end{proof}

\section{Auction Model for Bidding Clubs}\label{biddingclubs}


In this section we extend both the economic environment and auction mechanism from section \ref{auctionmodel} to include the characteristics necessary for a model of bidding clubs. Because our aim is not to model a situation where agents' \emph{decision} to collude is exogenous---as this would gloss over the question of whether the collusion is stable---we include the collusive protocol as part of the model and show that it is individually rational \emph{ex post} (i.e., after agents have observed their valuations) for agents to choose to collude.  However, we do consider exogenous the selection of the set of agents who are \emph{offered} the opportunity to collude. Furthermore, we want to show the impact of the possibility of collusion upon non-colluding agents; indeed, even colluding agents must take into account the possibility that \emph{other} groups of agents in the auction may also be colluding.  Once we have defined the new economic environment and auction mechanism, a well-defined Bayesian game will be specified by every tuple of primary auction type, bidding club rules and distributions of agent types, the number of agents and the number of bidding clubs.

\subsection{The Economic Environment}\label{bcenvironment}

We extend the economic environment $E$ from the previous section to consist of a set of agents who have non-negative valuations for a good at auction, the distinguished agent $0$ and a set of bidding club coordinators who may invite agents to participate in a bidding club. Intuitively, we construct an environment where an agent's belief update after observing the number of agents in his bidding club does not result in any change in the distribution over the number of \emph{other} agents in the auction, because the number of agents in each bidding club is independent of the number of agents in every other bidding club.

\subsubsection{Coordinators}

Coordinators are not free to choose their own strategies; rather, they act as part of the mechanism for a subset of the agents in the economic environment.  We select coordinators in a process analogous to our previous approach for exogenously selecting agents: we draw a finite set of individuals from an infinite set of potential coordinators. In this case, however, this finite set is considered ``potential coordinators''; in section \ref{agents} we will describe which potential coordinators are ``actualized'', i.e., correspond to actual coordinators.  Possible coordinators that are not actualized will correspond to singleton bidders in the auction.

More formally, let $\mathcal{C} \equiv \mathbb{N}$ (excluding 0) be the set of all coordinators. $\beta_C$ represents the probability that a finite set $C \subset \mathcal{C}$ is selected to be the set of potential coordinators. We add the restriction that all coordinators are equally likely to be chosen.  A consequence of this restriction is that an agent's knowledge of the coordinator with whom he is associated does not give him additional information about what other coordinators may have been selected. We denote the probability that an auction will involve $n_c$ potential coordinators as $\gamma_C(n_c) = \sum_{C, |C| = n_c} \beta_C$.  The distribution $\beta_C$ is common knowledge.  We assume that $\gamma_C(0) = \gamma_C(1) = 0$: at least two potential coordinators will be associated with each auction. 

\subsubsection{Agents}\label{agents}

We independently associate a random number of agents with each potential coordinator, again drawing a finite set of actual agents from an infinite set of potential agents.  If only one (actual) agent is associated with a potential coordinator, the potential coordinator will not be actualized and hence the agent will not belong to a bidding club.  In this way we model agents who participate directly in the auction without being associated with a coordinator. If more than one agent is associated with a potential coordinator, the coordinator \emph{is} actualized and all the agents receive an invitation to participate in the bidding club.


More formally, let $\mathcal{A} \equiv \mathbb{N}$ be the set of all agents, and let $\bcmax$ be the maximum number of agents who may be associated with a single bidding club.  Partition $\mathcal{A}$ into subsets, where agent $i$ belongs to the subset $\mathcal{A}_{\lceil i / \bcmax \rceil}$. Let $\beta_{A}$ be the probability that a finite set $A \subset \mathcal{A}_i$ is the set of agents associated with potential coordinator $i$; we assume that this distribution is the same for all $i$. Furthermore, as above, we assume that it is common knowledge that all agents are equally likely to be chosen.  The probability that $n$ agents will be associated with a potential coordinator is denoted $\gamma_A(n) = \sum_{A, |A| = n} \beta_A$.  By the definition of $\bcmax$, $\forall j > \bcmax, \gamma_A(j) = 0$; we assume that $\gamma_A(0) = 0$ and that $\gamma_A(1) < 1$.

\subsubsection{Signals}


Each agent receives a signal informing him of the number of agents in his bidding club; as above we denote this signal as $\signal_i$.\footnote{In fact, none of our results require that agents know the \emph{number} of agents in their bidding clubs; it would be sufficient that agents know \emph{whether} they belong to a bidding club.  We consider the setting where agents' signals are more informative because it simplifies the exposition of the main theorem.}  Of course, if this number is $1$ then there is no coordinator for the agent to deal with, and he will simply participate in the main auction. Note also that agents are neither aware of the number of potential coordinators for their auction nor the number of actualized potential coordinators, though they are aware of both distributions. 

\subsubsection{Beliefs}\label{beliefs}

Once an agent is selected, he updates his probability distribution over the number of actual agents in the economic environment. Not all agents will have the same beliefs---agents who have been signaled that they belong to a bidding club will expect a larger number of agents than singleton agents. We denote by $p_m^{n,k}$ the probability that there are a total of $m$ agents in the auction, given that there are $n$ bidding clubs and that there are $k$ agents in the bidder's own club; we denote the whole distribution $P^{n,k}$.  Because the numbers of agents in each bidding club are independent, observe that every agent in the whole auction has the same beliefs about the number of other agents in the economic environment, discounting those agents in his own bidding club.  Hence agent $i$'s beliefs are described by the distribution $P^{n,\signal_i}$. It is important to note that $P^{n,\signal_i}$ is simply another distribution over the number of agents in the auction. Although this shorthand makes reference to the bidding club economic environment in order to describe the construction of the distribution, it makes sense to talk about a classical auction with a stochastic number of bidders (i.e., section \ref{first-price-stochastic}) where the number of bidders is distributed according to $P^{n,k}$ for given values of $n$ and $k$.

\subsection{The Augmented Auction Mechanism}

Bidding clubs, in combination with a main auction, induce an augmented auction mechanism for their members:

\begin{enumerate}

\item A set $A$ of bidders is invited to join the bidding club. 

\item Each agent $i$ sends a message $\mu_i$ to the bidding club coordinator.  This may be the null message, which indicates that the agent will not participate in the coordination and will instead participate freely in the main auction. Otherwise, agent $i$ agrees to be bound by the bidding club rules, and $\mu_i$ is agent $i$'s declared valuation for the good. Of course, $i$ can lie about his valuation.

\item Based on pre-specified and commonly-known rules, and on the information all the members supply, the coordinator selects a subset of the agents to bid in the main auction. The coordinator may bid on behalf of these agents (e.g., using their ID's on the auction web site) or it may instruct agents on how to bid.  In either case we assume that the coordinator can force agents to bid as desired, for example by imposing a charge on agents who do not behave as directed.  

\item If a bidder represented by the coordinator wins the main auction, he is made to pay the amount required by the auction mechanism to the auctioneer. In addition, he may be required to make an additional payment to the coordinator.  



\end{enumerate}

Any number of coordinators may participate in an auction.  However, we assume that there is only a single coordination protocol, and that this protocol is common knowledge.  

\section{Bidding Clubs for First-Price Auctions}\label{first-price}

In this section we first give some (mild) assumptions about the distribution of agent valuations, then use these assumptions to prove a technical lemma.  We then give the bidding club protocol for first-price auctions.  We consider a first-price auction with participation revelation as described in section \ref{participationrevelation}.  Bidders indicate their intention to participate, the auctioneer announces the total number of bidders and then bidders place their bids.  The bidding club decides whether to drop bidders before the first phase; therefore the number announced by the auctioneer does not include dropped bidders.  We show an equilibrium of this auction, and demonstrate that agents gain under this equilibrium.

\subsection{Assumptions}\label{assumption}

Our results hold for a broad class of distributions of agent valuations---all distributions for which the following two assumptions are true.

First, we assume that $F$ is continuous and atomless.  

In order to give our second assumption, we must introduce some notation.  Define:

\begin{equation} 
P_{x \geq i} = \sum_{x=i}^{\infty} p_x. 
\end{equation}

We now define the relation ``$<$'' for probability distributions:

\begin{equation}
P < P' \, \textrm{ iff } \, \exists l ( \forall i < l,  P_{x \geq i}=p'_{x \geq i} \textrm{ and } \forall i \geq l,  P_{x \geq i} < P'_{x \geq i}).\label{geq}
\end{equation}

We are now able to state our second assumption:

\begin{equation}
(P < P') \textrm{ implies that } \forall v, \, b^e(v,P) < b^e(v,P'),
\end{equation}

Intuitively, we assume that every agent's symmetric equilibrium bid in a setting with a stochastic number of participants drawn from $P'$ is strictly greater than that agent's symmetric equilibrium bid in a setting with a stochastic number of participants drawn from $P$, in the case where $P'$ stochastically dominates $P$.


\subsection{A Technical Lemma}


Recall from section \ref{beliefs} that the notation $P^{n,k}$ may be seen as defining a probability distribution over the number of agents that is independent of the bidding club setting. It is thus possible to discuss equilibrium bids in the classical stochastic settings where the number of bidders is drawn from such a distribution.  While it will remain to show why these values are meaningful in our setting where (among other differences) agents have asymmetric information, it will be useful to prove the following lemma about the classical stochastic setting:

\begin{lemma} \label{assumptionlemma} $\forall k\geq2, \forall n \geq 2, \forall v, \, b^e(v,P^{n+k-1,1}) > b^e(v,P^{n,k})$ \end{lemma}

\emph{Remark. }  For convenience and to preserve intuition in what follows we will refer to the number of potential coordinators and the number of agents belonging to a coordinator even though we concern ourselves with the classical economic environment from section \ref{econenviron} where bidding clubs do not exist.  The number of potential coordinators is shorthand for the number $n_c$ drawn from $\gamma_C$ in the first phase of the procedural definition of the distribution $P^{n,k}$.  Likewise the number of agents associated with a potential coordinator is shorthand for the number of agents chosen from one of the $n_c$ iterative draws from $\gamma_A$. Intuitively, this lemma asserts that the symmetric equilibrium bid is always higher when more agents belong to the main auction as singleton bidders and the total number of agents is held constant.  

\begin{proof}Recall our second assumption from section \ref{assumption}.  We defined $P < P'$ as the proposition that $\exists l ( \forall i < l,  P_{x \geq i}=P'_{x \geq i} \textrm{ and } \forall i \geq l,  P_{x \geq i} < P'_{x \geq i})$.  Our second assumption was that $(P < P')$ implies that $\forall v, b^e(v,P) < b^e(v,P')$. It is thus sufficient to show that $P^{n+k-1,1} > P^{n,k}$.  We will take $l = n+k$.  

First we will show that $\forall j < n+k, P^{n+k-1,1}_{x\geq j} = P^{n,k}_{x\geq j}$.  The distribution $P^{n+k-1,1}$ expresses the belief that there are $n+k-2$ potential coordinators, the membership of which is distributed as described in section \ref{bcenvironment}, and one potential coordinator that is known to contain only a single bidder.  The distribution $P^{n,k}$ expresses the belief that there are $n-1$ potential coordinators, the membership of which is again distributed as described in section \ref{bcenvironment}, and one potential coordinator that is known to contain exactly $k$ bidders. Under both distributions it is certain that there are at least $n+k-1$ agents.  Therefore $\forall j < n+k, P^{n+k-1,1}_{x\geq j} = P^{n,k}_{x\geq j} = 1$.

Second, $\forall j \geq n+k$, $P^{n+k-1,1}_{x\geq j} > P^{n,k}_{x\geq j}$.  Considering $P^{n+k-1,1}$, observe that for $n+k-2$ of the potential coordinators the probability that this coordinator contains a single agent is less than one and these probabilities are all independent; the last potential coordinator contains a single agent with probability one.  Considering $P^{n,k}$, there are $n-1$ potential coordinators where the probability of containing a single agent is less than one, exactly as above, and $k$ potential coordinators certain to contain exactly one agent.  Thus the two distributions agree exactly about $n-1$ of the potential coordinators, which both hold to contain more than a single agent, and likewise both distributions agree that one of the potential coordinators contains exactly one agent. However, there remain $k-1$ potential coordinators about which the distributions disagree; $P^{n+k-1,1}$ always generates a greater or equal number of agents for these potential coordinators, as compared to $P^{n,k}$.  Under the latter distribution all these agents are singletons with probability one, while under the former there is positive probability that each of the potential coordinators contains more than one agent. As long as $k\geq 2$, there is at least one potential coordinator for which $P^{n+k-1,1}$ stochastically dominates $P^{n,k}$. Thus $\forall k\geq2, \forall n \geq 2, \forall v \, P^{n+k-1,1} >P^{n,k}$.
\end{proof}



\subsection{First-Price Auction Bidding Club Protocol}

What follows is the protocol of a coordinator who approaches $k$ agents.  

\begin{enumerate} 
\item Each agent $i$ sends a message $\mu_i$ to the coordinator.

\item If at least one agent declines participation then the coordinator registers in the main auction for every agent who accepted the invitation to the bidding club.  For each bidder $i$, the coordinator submits a bid of $b^e(\mu_i, P^{n,k})$, where $n$ is the number of bidders announced by the auctioneer.


\item If all $k$ agents accepted the invitation then the coordinator drops all bidders except the bidder with the highest reported valuation, who we will denote as bidder $h$.  For this bidder the coordinator will place a bid of $b^e(\mu_h,P^{n,1})$ in the main auction.


\item If bidder $h$ wins in the main auction, he is made to pay $b^e(\mu_h,P^{n,1})$ to the center and $b^e(\mu_h,P^{n,k}) - b^e(\mu_h,P^{n,1})$ to the coordinator.



\end{enumerate} 

We are now ready to prove the main theorem of the paper:

\begin{theorem}It is an equilibrium for all bidding club members to choose to participate and to truthfully declare their valuations to their respective bidding club coordinators, and for all non-bidding club members to participate in the main auction with a bid of $b^e(v,P^{n,1})$.\label{theorem-firstprice}\end{theorem}

\begin{proof} We first prove that the above strategy is in equilibrium for both categories of bidders given that agents all participate; we then prove that participation is rational for all agents.  

For the proof of equilibrium we consider a one-stage mechanism which behaves as follows: 

\begin{enumerate}
\item The center announces $n$, the number of bidders in the main auction.
\item Bidders submit bids (messages) to the mechanism.
\item The bidder with the highest bid is allocated the good.
\item The winning bidder is made to pay $b^e(v_i,P^{n,\signal_i})$.
\end{enumerate}

This one-stage mechanism has the same payment rule for bidding club bidders as the bidding club protocol given above, but no longer implements a first-price payment rule for singleton bidders.  In order to prove that the strategies given in the statement of the theorem are an equilibrium, it is sufficient to show that truthful bidding is an equilibrium for all bidders under the one-stage mechanism.  Observe that this mechanism may be seen as a mechanism $\bar{M}$ in the sense of lemma \ref{lemma}: it allocates the good to the agent who submits the highest message, and (by definition of $b^e$) the auction $M_i$ in which \emph{all} agents are subject to agent $i$'s payment rule and receive the signal $\signal_i$ has truth revelation as a symmetric equilibrium. 

\emph{Strategy of non-club bidder: } Assume that all bidding club agents bid truthfully.  Further assume that all non-club agents also bid truthfully except for agent $i$. The probability distribution $P^{n,1}$ correctly describes the beliefs of non-club agents, given the auctioneer's announcement that there are $n$ bidders in the main auction.  Although agents in bidding clubs have additional information about the number of agents---each agent knows that there is at least one other agent in his own club---their prescribed behavior is to place bids of $b^e(\mu,P^{n,1})$ in the main auction.  Agent $i$ thus faces a stochastic number of agents distributed according to $P^{n,1}$ and all bidding $b^e(v,P^{n,1})$.  Using the result from lemma \ref{lemma}, $i$'s strategic decision is the same as under a mechanism where all agents are subject to his payment rule and share his signal $\signal_i$, and with a stochastic number of bidders distributed according to $P^{n,1}$.  In particular, it does not matter that the club members are subject to different payment rules and have additional information, and so $i$ will also bid $b^e(v,P^{n,1})$.

\emph{Strategy of club bidder: } Assume that all agents accept the invitation to join their respective clubs and then truthfully declare their valuations, excluding agent $i$ who decides to participate but considers his bid. Once again, observe that $i$ is in a setting that is exactly described by lemma \ref{lemma}: $P^{n,k}$ really does describe the distribution over the number of agents given his signal, and the bidder submitting the highest (global) message will always be allocated the good.  Therefore the information asymmetry does not affect $i$'s strategy, and so truthful bidding is a best response for agent $i$.

We now turn to the question of participation; for this part of the proof we return to the original, multi-stage mechanism.

\emph{Participation of non-club bidder: } Because there is no participation fee, it is always rational for a bidder to participate in a first-price auction.

\emph{Participation of club bidder: } Likewise, because there is no participation fee, all bidding club bidders will participate in the auction, but must decide whether or not to accept their coordinators' invitations. Assume that all agents except for $i$ join their respective clubs and bid truthfully, and agent $i$ must decide whether or not to join his bidding club.  Agent $i$ knows the number of agents in his bidding club and updates his distribution over the number of agents in the whole auction as $P^{n,k}$.  

Consider the classical stochastic case where all bidders have the same information as $i$ (and are subject to the same payment rules): from proposition \ref{propstochastic} it is a best response for $i$ to bid $b^e(v_i,P^{n,k})$. In this setting $i$'s expected gain is the same as in the equilibrium where all bidding club members (including $i$) join their clubs and bid truthfully, by corollary \ref{asymm-cor}. 

As a result of $i$ declining the offer to participate in the bidding club there are $n-1$ bidders in the main auction placing bids of $b^e(v,P^{n+k-1,1})$ and $k-1$ other bidders placing bids of $b^e(v,P^{n,k})$.  Note that this occurs because the singleton bidders and other bidding clubs in the main auction follow a strategy that depends on the number of bidders announced by the auctioneer; hence they bid as though all the $k-1$ bidders from the disbanded bidding club might each be independent bidding clubs. We know from lemma \ref{assumptionlemma} that $b^e(v,P^{n+k-1,1}) > b^e(v,P^{n,k})$.  Thus the singleton bidders and other bidding clubs will bid a higher function of their valuations than the bidders from the disbanded bidding club.  It always reduces a bidder's expected gain in a first-price auction to cause other bidders to bid above the equilibrium, because it reduces the chance that he will win without affecting his payment if he does win.  This is exactly the effect of $i$ declining the offer to join his bidding club: the $k-1$ other bidders from $i$'s bidding club bid according to the equilibrium of the classical stochastic case discussed above, but the $n-1$ singleton and bidding club bidders submit bids that exceed the symmetric equilibrium amount. Therefore $i$'s expected gain is smaller if he declines the offer to participate than if he accepts it.
\end{proof}

\subsection{Do bidding clubs cause agents to gain?}

We can show that bidders are better off being invited to a bidding club than being sent to the auction as singleton bidders. Intuitively, an agent gains by not having to consider the possibility that other bidders who would otherwise have belonged to his bidding club might themselves be bidding clubs.  

\begin{theorem}\label{theorem-gain-disband}
An agent $i$ has higher expected utility in a bidding club of size $k$ bidding as described in theorem \ref{theorem-firstprice} than he does if the bidding club does not exist and $k$ additional agents (including $i$) participate directly in the main auction as singleton bidders, again bidding as described in theorem \ref{theorem-firstprice}.
\end{theorem}


\begin{proof} 
Consider the counterfactual case where agent $i$'s bidding club does not exist, and all the members of this bidding club become singleton bidders.  We will show that $i$ is better off as a member of the bidding club than in this case.  If there were $n$ potential coordinators in the original auction and $k$ agents in $i$'s bidding club, then the auctioneer will announce $n+k-1$ as the number of participants in the new auction. Under the equilibrium from theorem \ref{theorem-firstprice}, as a singleton bidder $i$ will bid $b^e(v_i,P^{n+k-1,1})$.  If he belonged to the bidding club and followed the same equilibrium $i$ would bid $b^e(v_i,P^{n,k})$.  In both cases the auction is economically efficient, which means $i$ is better off in the auction that requires him to pay a smaller amount when he wins.  Lemma \ref{assumptionlemma} shows that $\forall k\geq 2, \forall n \geq 2, \forall v,  b^e(v,P^{n+k-1,1}) > b^e(v,P^{n,k})$, and so our result follows.
\end{proof}

We can also show that singleton bidders and members of other bidding clubs benefit from the existence of each bidding club in the same sense.  Following an argument similar to the one in theorem \ref{theorem-gain-disband}, other bidders gain from not having to consider the possibility that additional bidders might represent bidding clubs.  Paradoxically, other bidders' gain from the existence of a given bidding club is greater than the gain of that club's members.

\begin{corollary}
In the equilibrium described in theorem \ref{theorem-firstprice}, singleton bidders and members of other bidding clubs have higher expected utility when other agents participate in a given bidding club of size $k \geq 2$, as compared to a case where $k$ additional agents participate directly in the main auction as singleton bidders.
\end{corollary}

\begin{proof}
Consider a singleton bidder in the first case, where the club of $k$ agents does exist. (It is sufficient to consider singleton bidders, since other bidding clubs bid in the same way as singleton bidders.) Following the equilibrium from theorem \ref{theorem-firstprice} this agent would submit the bid $b^e(v_i,P^{n,1})$.  Theorem \ref{theorem-gain-disband} shows that it is better to belong to a bidding club (and thus to bid $b^e(v_i,P^{n,k})$) than to be a singleton bidder in an auction with the same number of agents (and thus to bid $b^e(v_i,P^{n+k-1,1})$.  Since the distribution $P^{n,k}$ is just $P^{n,1}$ with $k-1$ singleton agents added, $\forall k \geq 2, b^e(v_i,P^{n,1}) < b^e(v_i,P^{n,k})$.  Thus $\forall k \geq 2, b^e(v_i,P^{n,1}) < b^e(v_i,P^{n+k-1,1})$.
\end{proof}

Finally, we can show that agents are indifferent between participating in the equilibrium from theorem \ref{theorem-firstprice} in a bidding club of size $k$ (thus, where the number of agents is distributed according to $P^{n,k}$) and participating in an economic environment with a stochastic number of bidders distributed according to $P^{n,k}$, but with no coordinators.

\begin{theorem}For all $\type_i \in \mathcal{T}$, for all $k \geq 1$, for all $n \geq 2$, agent $i$ obtains the same expected utility by:
\begin{enumerate}
\item participating in a bidding club of size $k$ in the economic environment from section \ref{bcenvironment} and following the equilibrium from theorem \ref{theorem-firstprice};
\item participating in a first-price auction with participation revelation in an economic environment with a stochastic number of bidders distributed according to $P^{n,k}$ where all bidders receive the null signal, and where there are no coordinators.
\end{enumerate}
\end{theorem}

\begin{proof}
First we will show that agent $i$'s expected utility in case (2) above is the same as in a classical first-price auction with a stochastic number of bidders (i.e., without participation revelation).  Second, we will show that agent $i$'s expected utility in this classical stochastic setting is the same as in case (1) above.

From proposition \ref{prop-participationrevelation} it is an equilibrium for agent $i$ to bid $b^e(v_i,j)$ in a first-price auction with participation revelation (case (2)), where $j$ is the number of bidders announced by the auctioneer.  Since the number of agents is distributed according to $P^{n,k}$, the expected payment of agent $i$ is $\sum_{j=2}^{\infty} p^{n,k}_j b^e(v_i,j)$.  This is the definition of $b^e(v_i,P^{n,k})$ from equation \ref{stochasticbid}. From proposition \ref{propstochastic} this is an equilibrium bid of agent $i$ when the number of agents is distributed according to $P^{n,k}$ (without information revelation).  Since both the classical first-price auction with a stochastic number of bidders and the first-price auction with participation revelation are efficient, agent $i$'s expected utility is the same under both auctions.

Under the equilibrium from theorem \ref{theorem-firstprice} (case (1)) the amount of $i$'s payment will be $b^e(v_i,P^{n,k})$ if he wins.  Since both the mechanism from case (1) and the classical first-price auction with a stochastic number of bidders are efficient, agent $i$ has the same expected utility in both auctions. 
\end{proof}

This theorem shows that an agent would be as happy in a world without bidding clubs as he is in our economic environment.  The difference between the two worlds is that in the latter bidding club coordinators make a positive profit on expectation, and indeed never lose money.  That is, in the bidding club economic environment some expected profit is shifted from the auctioneer to the bidding club coordinator(s) without affecting the bidders' expected utility. We observe that it would be easy for coordinators to redistribute some of these gains to bidders along the lines of the second-price auction protocol proposed by Graham and Marshall: coordinators make a payment to every bidder who accepts the invitation to join, where the amount of this payment is less than or equal to the \emph{ex ante} expected difference that bidder makes to the coordinator's profit.  With this modification coordinators would be budget balanced only on expectation (violating requirement \ref{makemoney} from section \ref{bc-glance}), but agents would strictly prefer the bidding club economic environment to the economic environment in which coordinators are not present.



\section{Discussion}\label{discussion}

In this section we consider the trustworthiness and legality of coordinators, and also discuss two ways for auctioneers to disrupt bidding clubs in their auctions.

\subsection{Trust}

Why would a bidding club coordinator be willing to provide reliable service, and likewise why would bidders have reason to trust a coordinator?  For example, a malicious coordination protocol could be used simply to drop all its members from the auction and reduce competition.  While this is a reasonable concern, all the bidding club protocols discussed in this paper allow the coordinator to make a profit on expectation.  There is thus incentive for a trusted third party to run a reliable coordination service.  Indeed, coordinators would be very inexpensive to run: as their behavior is entirely specified, they could operate without any human supervision. The establishment of trust is exogenous to our model; we have simply assumed that all agents trust coordinators and that all coordinators are honest.

\subsection{Legality}

We have often been asked about the legal issues surrounding the use of bidding clubs.  While this is an interesting and pertinent question, it exceeds both our expertise and the scope of this paper.  We should note, however, that uses of bidding clubs exist that might not fall under the legal definition of collusion.  For example, a corporation could use a bidding club to choose one of its departments to bid in an external auction.  In this way the corporation could be sure to avoid bidding against itself in the external auction while avoiding dictatorship and respecting each department's self-interest.  Coordinators may also be permitted by the auctioneer: e.g., by an internet market seeking to attract more bidders to its site.

\subsection{Disrupting Bidding Clubs}


There are two things an auctioneer can do to disrupt bidding clubs in a first-price auction.  First, she can permit ``false-name bidding.''  Our auction model has assumed that each agent may place only a single bid in the auction, and that the center has a way of uniquely identifying agents.  For example, the auctioneer might use user accounts keyed to credit card billing addresses in combination with a reputation ranking, making it impossible for bidders to place bids claiming to originate from different agents.  Second, she can refrain from publicly disclosing the winner of the auction.

If bidders can bid both in their bidding clubs and in the main auction, they are better off deviating from the equilibrium in theorem \ref{theorem-firstprice} in the following way.  A bidder $i$ can accept the invitation to join the bidding club but place a very low bid with the coordinator; at the same time, $i$ can directly submit a competitive bid in the main auction.  Agent $i$ will gain by following this strategy when all other agents follow the strategies specified in theorem \ref{theorem-firstprice} because accepting the invitation to join the bidding club ensures that the club does drop all but one of its members and also causes the high bidder to bid less than he would if he were not bound to the coordination protocol.  If the bidding club drops any bidders other than $i$ then all agents' bids will also be lowered because the number of participants announced by the auctioneer will be smaller, compared to the case where the bidding club did not exist or where it was disbanded. However, if false-name bidding is impossible and the winner of the auction is publicly disclosed then the bidding club coordinator can detect an agent who has deviated in this way. Because the agent has agreed to participate in the bidding club the coordinator has the power to impose a punitive fine on this agent, making the deviation unprofitable.  If either or both of these requirements does not hold, however, the coordinator will be unable to detect defection and so the equilibrium from theorem \ref{theorem-firstprice} will not hold.

\section{Conclusion}\label{conclusion}

We have presented a formal model of bidding clubs which departs in many ways from models traditionally used in the study of collusion; most importantly, all agents behave strategically based on correct information about the economic environment, including the possibility that other agents will collude.  Other features of our setting include a stochastic number of agents and a stochastic number of bidding clubs in each auction.  Agents' strategy space is expanded so that the decision of whether or not to join a bidding club is part of an agent's choice of strategy. Bidding clubs never lose money, and gain on expectation.  We have showed a bidding club protocol for first-price auctions that leads to a (globally) efficient allocation in equilibrium, and which does not make use of side-payments.  There are three ways of asking the question of whether agents gain by participating in bidding clubs in first-price auctions: 

\begin{enumerate}
\item Could any agent gain by deviating from the protocol?
\item Would any agent be better off if his bidding club did not exist?
\item Would any agent would be better off in an economic environment that did not include bidding clubs at all?
\end{enumerate}  

We have showed that agents are strictly better off in the first two senses and no worse off in the last sense; furthermore, we have described a simple side-payment scheme that would make agents strictly better off in all three senses.  We have also showed that each bidding club causes \emph{non-members} to gain in the second sense.  Finally, we have discussed ways for an auctioneer to set up the rules of her auction so as to disrupt bidding clubs.


\begin{thebibliography}{}

\bibitem[Cramton and Palfrey, 1990]{CramtonPalfrey}
Cramton, P. and Palfrey, T. (1990).
\newblock Cartel enforcement with uncertainty about costs.
\newblock {\em International Economic Review}, 31(1):17--47.

\bibitem[Feinstein et~al., 1985]{FeinsteinBlockNold}
Feinstein, J., Block, M., and Nold, F. (1985).
\newblock Assymetric behavior and collusive behavior in auction markets.
\newblock {\em The American Economic Review}, 75(3):441--460.

\bibitem[Graham and Marshall, 1987]{GrahamMarshall}
Graham, D. and Marshall, R. (1987).
\newblock Collusive bidder behavior at single-object second-price and english
  auctions.
\newblock {\em Journal of Political Economy}, 95:579--599.

\bibitem[Graham et~al., 1990]{GrahamMarshallRichard}
Graham, D., Marshall, R., and Richard, J.-F. (1990).
\newblock Differential payments within a bidder coalition and the shapley
  value.
\newblock {\em The American Economic Review}, 80(3):493--510.

\bibitem[Harstad et~al., 1990]{HarstadKagelLevin}
Harstad, R., Kagel, J., and Levin, D. (1990).
\newblock Equilibrium bid functions for auctions with an uncertain number of
  bidders.
\newblock {\em Economic Letters}, 33(1):35--40.

\bibitem[Hendricks and Porter, 1989]{HendricksPorter}
Hendricks, K. and Porter, R. (1989).
\newblock Collusion in auctions.
\newblock {\em Annale's D'economie de Statistique}, 15/16:216--229.

\bibitem[Leyton-Brown et~al., 2000]{biddingclubs}
Leyton-Brown, K., Shoham, Y., and Tennenholtz, M. (2000).
\newblock Bidding clubs: institutionalized collusion in auctions.
\newblock In {\em ACM Conference on Electronic Commerce}.

\bibitem[Mailath and Zemsky, 1991]{MailathZemsky}
Mailath, G. and Zemsky, P. (1991).
\newblock Collusion in second-price auctions with heterogeneous bidders.
\newblock {\em Games and Economic Behavior}, 3:467--486.

\bibitem[McAfee and McMillan, 1987]{McAfeeRandom}
McAfee, R. and McMillan, J. (1987).
\newblock Auctions with a stochastic number of bidders.
\newblock {\em Journal of Economic Theory}, 43:1--19.

\bibitem[McAfee and McMillan, 1992]{McAfeeMcMillan92}
McAfee, R. and McMillan, J. (1992).
\newblock Bidding rings.
\newblock {\em The American Economic Theory}, 82:579--599.

\bibitem[Monderer and Tennenholtz, 2000]{MonTenaij}
Monderer, D. and Tennenholtz, M. (2000).
\newblock {Optimal Auctions Revisited}.
\newblock {\em Artificial Intelligence}, 120(1):29--42.

\bibitem[Riley and Samuelson, 1981]{RileySamuelson}
Riley, J. and Samuelson, W. (1981).
\newblock Optimal auctions.
\newblock {\em American Economic Review}, 71:381--392.

\bibitem[Robinson, 1985]{Robinsoncoll}
Robinson, M. (1985).
\newblock Collusion and the choice of auction.
\newblock {\em Rand Journal of Economics}, 16(1):141--145.

\bibitem[von Ungern-Sternberg, 1988]{Ungern-Sternberg}
von Ungern-Sternberg, T. (1988).
\newblock Cartel stability in sealed bid second price auctions.
\newblock {\em The Journal of Industrial Economics}, 18(3):351--358.

\end{thebibliography}

\end{document}